# ATTITUDE CONTROL OF AN INFLATABLE SAILPLANE FOR MARS EXPLORATION

Adrien Bouskela,[*] Aman Chandra,[†]
Jekanthan Thangavelautham,[‡] Sergey Shkarayev[§]

Exploration of Mars has been made possible using a series of landers, rovers and orbiters. The HiRise camera on the Mars Reconnaissance Orbiter (MRO) has captured high-resolution images covering large tracts of the surface. However, orbital images lack the depth and rich detail obtained from in-situ exploration. Rovers such as Mars Science Laboratory and upcoming Mars 2020 carry state-of-the-art science laboratories to perform in-situ exploration and analysis. However, they can only cover a small area of Mars through the course of their mission. A critical capability gap exists in our ability to image, provide services and explore large tracts of the surface of Mars required for enabling a future human mission. A promising solution is to develop a reconnaissance sailplane that travels tens to hundreds of kilometers per sol. The aircraft would be equipped with imagers that provide that in-situ depth of field, with coverage comparable to orbital assets such as MRO. A major challenge is that the Martian carbon dioxide atmosphere is thin, with a pressure of 1% of Earth at sea level. To compensate, the aircraft needs to fly at high-velocities and have sufficiently large wing area to generate the required lift. Inflatable wings are an excellent choice as they have the lowest mass and can be used to change shape (morph) depending on aerodynamic or control requirements. In this paper, we present our design of an inflatable sailplane capable of deploying from a 12U CubeSat platform. A pneumatic deployment mechanism ensures highly compact stowage volumes and minimizes complexity. The present work attempts to describe expected dynamic behavior of the design and contributes to evolving an effective strategy for attitude control required for stable flight and high-quality imaging. The use of Dynamic Soaring as a means of sustained unpowered flight in the low-density Martian atmosphere will be studied through a point mass sailplane model. Using a linear wind gradient model of the Martian atmospheric boundary layer, numerical simulations of such trajectories will attempt to demonstrate that longer duration missions can be conducted using such hardware and flight characteristics.

## INTRODUCTION

Exploration of Mars has been made possible with a series of orbiters, landers and rovers that have provided unprecedented view of the Red Planet's surface. Orbiters such as Mars Reconnaissance Orbiter (MRO) are equipped with the HiRise camera that provides a resolution of 0.3 m/pixel[1].

---

[*] Graduate Student first author, Aerospace & Mechanical Engineering, Univ. of Arizona, 1130 N Mountain, Tucson AZ.
[†] Graduate Student first author, Aerospace & Mechanical Engineering, Univ. of Arizona, 1130 N Mountain, Tucson AZ.
[‡] Assistant Professor, Aerospace & Mechanical Engineering, Univ. of Arizona, 1130 N Mountain Ave, Tucson AZ.
[§] Professor, Aerospace & Mechanical Engineering, Univ. of Arizona, 1130 N Mountain Ave, Tucson AZ.

Landers including Mars Phoenix and Insight can achieve ground resolutions of 0.01 m/pixel or higher but are restricted to imaging a region within a 100 m diameter. Rovers such as the Mars Science Laboratory (MSL) and the upcoming Mars 2020 are equipped with state-of-the-art laboratories to perform in-situ experiments. They can drive up to several hundred meters a day. Due to limitations with current Mars Entry Descent and Landing (EDL) technology current Mars surface assets have landed in regions that are relatively benign.

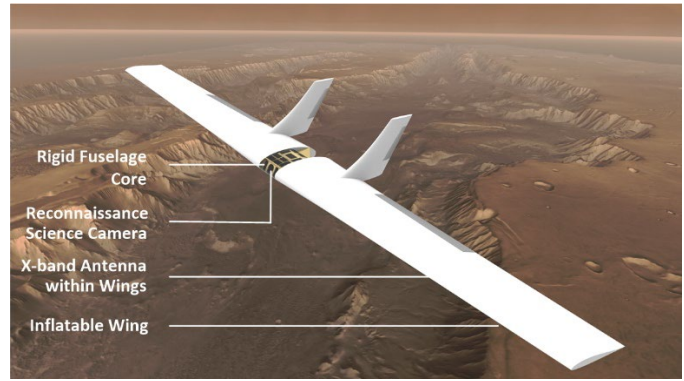

**Figure 1.** An Inflatable Sailplane Aircraft for Science Reconnaissance on Mars.

The emerging science questions about Mars, such as past geo-history, past-atmosphere and past habitability all require access to more rugged surfaces, including the highlands, crater walls, canyons and cliffs. These surface features are often located hundreds of kilometers from a landing site. Furthermore, traversal through this rugged environment by ground may take years. A better solution is needed to perform continental-scale science reconnaissance of Mars.

A Mars aircraft equipped with reconnaissance cameras can address the emerging limitations of Mars orbital assets and landers/rovers. We propose a radically different approach by deploying a 12U, 24 kg CubeSat that transforms into a 5 kg inflatable Mars sailplane aircraft (Figure 1) that will perform science-led surface reconnaissance of Mars with image resolutions of 0.1 m/pixel to 0.01 m/pixel and fly at 11 m/s to 100 m/s. This is 3 to 30 times higher resolution than the state-of-the-art HiRISE camera onboard Mars Reconnaissance Orbiter[1,2].

 The 24-kg CubeSat package would utilize some of the 190 kg of ballast tungsten available on the flagship Mars Science Laboratory and upcoming Mars 2020. It would be dropped off during EDL, when the main payload has slowed down to 100 m/s and use its onboard propulsion system to achieve a 1 km separation distance from the rover. The proposed Mars sailplane defines a new architecture to enable missions in the Mars atmosphere at 1/10 cost, 1/50 mass and volume of previous missions. The Mars sailplane can provide an important support role for high-priority Decadal Science such as Mars sample return[3]. This is going to be made possible through the CubeSat revolution[4] and the recent successful MarCO CubeSats[5,6].

The CubeSat package will deploy an inflatable wing that will keep the sailplane aloft. Inflatable wings are not new and have been demonstrated on Earth by Goodyear[34]. The key to staying aloft will be to measure atmospheric conditions of Mars and perform Dynamic Soaring[27,39,40,42,43]. This is a technique used by sailplane pilots to stay aloft for prolonged periods in the atmosphere without any engines. The rising airstreams are used to help the aircraft buildup potential energy which be converted to high velocity at low altitudes. The proposed CubeSat will have autonomous dynamic soaring capability enabling it to sample the neighborhood wind conditions and attain kinetic energy.

Among limitations and challenges of the previous research and development efforts, the short flight endurance is the most critical one. To overcome this underperformance, the unpowered sailplane will employ the dynamic soaring method for flight in the Martian atmosphere. The sailplane will detect and fly into a horizontal wind field which exhibits an acceptable vertical gradient in velocity magnitude. By doing so, it will accumulate additional kinetic and potential energy. This high-risk, high-reward mission has a potential to provide days of continues flight using advanced autonomous controls. Dynamic soaring has been proven successful in earth's atmosphere by sail-plane pilots and soaring birds. Numerical modeling will extend its validity to the Martian environment. To extend the range of sources of energy available to the sailplane the autonomous exploitation of thermal updrafts also known to exist on mars will also be explored.

In the following section we present previous related work followed by presentation of the sailplane concept. This will include overview of the sail plane design, deployment and plans for dynamic soaring in the Mars atmosphere.

**PREVIOUS RELATED WORK**

The proposed Mars sailplane can provide access to regions that were previously inaccessible. This will be possible through a high-risk, high-reward philosophy, but using proven low-cost components. There have been several Mars aircraft proposed including ARES[7](SCOUT program), Kitty Hawk, AME and MAGE (Discovery) [8]. Another is Dragonfly, an aircraft being proposed for Titan[9]. Another is the Mars Helicopter concept that is to fly alongside Mars 2020 rover[10]. These aircraft have mostly been designed to keep a science focus. The designed systems have all weighed over a 100 kg at a proposed a development cost above $350. These earlier aircraft concepts required a dedicated mission to Mars.

The previous aircraft concepts also needed an onboard propulsion system that range from turb-props suited for the Martian atmosphere to jet and rocket engines. These engines all required considerable fuel to stay-afloat and add to the mass and complexity of the aircraft. The resultant flight time was limited, with ARES baselined to achieve one hour of flight. With a sailplane concept, we remove the mass, volume and complexity of an engine, making the aircraft substantially lighter. The initial kinetic energy is provided during deployment during the Entry, Descent and Landing (EDL) sequence of the primary payload. This is well suited for the aircraft to exploit warm updrafts thermals and cold down-drafts. Such techniques have been utilized on earth and earth sailplanes have been demonstrated to stay flying for unlimited time periods. Independent of the technology development risks, the Martian atmosphere is known to vary throughout the seasons, increasing in density in certain seasons while decreasing in others. This presents additional challenges in the design, planning and longevity of the mission.

An alternative to flying is hopping. A typical approach to hopping is to use a hopping spring mechanism to overcome large obstacles[11]. One is the Micro-hopper for Mars exploration developed by the Canadian Space Agency[12]. The Micro-hopper has a regular tetrahedron geometry that enables it to land in any orientation at the end of a jump. The hopping mechanism is based on a novel cylindrical scissor mechanism enabled by a Shape Memory Alloy (SMA) actuator. However, the design allows only one jump per day on Mars.

Another technique for hopping developed by Plante and Dubowsky at MIT utilize Polymer Actuator Membranes (PAM) to load a spring. The system is only 18 grams and can enable hopping of Microbots with a mass of 100 g up to a 1 m[13,14]. Microbots are cm-scale spherical mobile robots equipped with power and communication systems, a mobility system that enables it to hop, roll and bounce and an array of miniaturized sensors such as imagers, spectrometers, and chemical analysis sensors developed at MIT[13,14]. They are intended to explore caves, lava-tubes, canyons and cliffs. Ideally, many hundreds of these robots would be deployed enabling large-scale in-situ exploration. Mapping and localization of cave environments using familiar techniques such as Simultaneous Localization and Mapping (SLAM) have been shown recently[15,16]. However, current techniques still don't account for the limited lighting conditions.

We are also developing the SphereX[17,18,19] robot that would perform long duration hops interspersed with short flights. SphereX is baselined to use a rocket engine and operate like a quadcopter in off-world environments. The platform has also been designed to use a mechanical hop-per[19]. Propulsion options considered include use of $CO_2$, water-steam, hydrogen-peroxide and kerosene. A simpler off-shot of the SphereX platform is the SPEER[20] platform which will used solid rockets to enter rugged environments and perform limited reconnaissance.

**MARS SAILPLANE CONCEPT**

The overall layout of the Mars Sailplane in its stowed and deployed state is shown in Figure 2 and Figure 3 respectively. The entire package fits into a 12U frame, with top 6U for the aircraft and bottom 6U for the gas generator (Table 1 and 2). The actual mass of the craft minus the cold gas propulsion is only 5 kg.

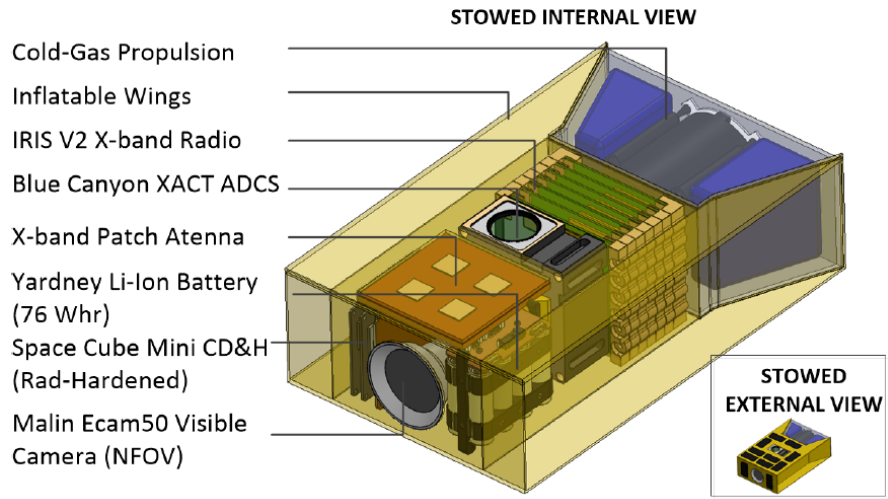

**Figure 2. Platform avionics and instrumentation in stowed form**

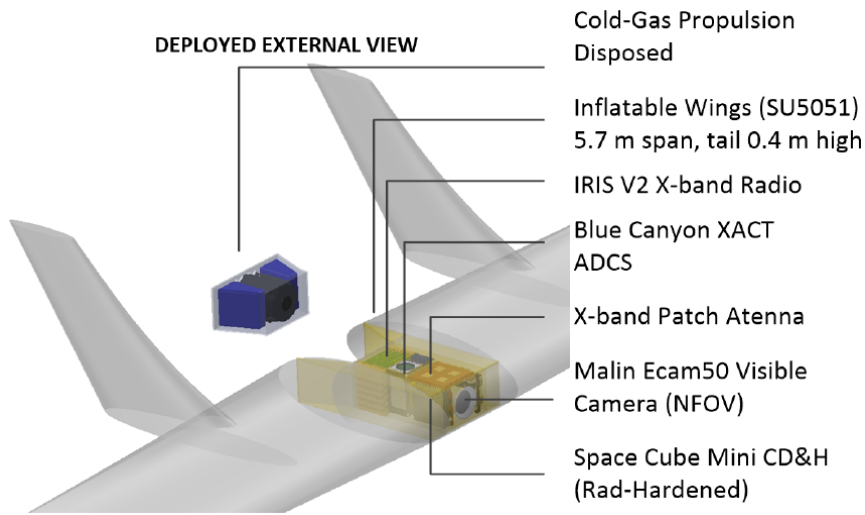

**Figure 3. Deployed inflatable control surfaces**

The sailplane is equipped with a Malin Ecam 50, 5 MP camera for science reconnaissance and for autonomous navigation. In addition, the craft is equipped with a rad-hardened Onboard Computer called the Space Cube Mini. For communications, the craft will use the IRIS X-band v2.1 radio to communicate with Mars relay assets such as the MRO. The onboard power system contains Yardney lithium-ion batteries that will be charged and topped up using body mounted solar panels.

The Martian atmosphere is very thin that its density ($\rho = 0.0137$ kg/m³) is a hundred times lower than that on Earth. Even though the gravity on the Mars is significantly lower, $g = 3.72$ kg m/s², it still makes an endeavor to achieve a sustained flight on this planet. Wing models of rectangular planform, with aspect ratio of 5 were tested[21] in the Reynolds number range from 20,000 to 170,000. A 12% maximum thickness wing, 2.9% thickness flat plate (0% camber), and cambered (5.8% camber) plate were investigated. It was found that the maximum lift coefficient of a thin, cambered-plate wing ($C_{Lmax} = 1.05$) was nearly twice that obtained for a traditional 12% thickness wing. Results of the previous studies [22,23,24] confirmed the superiority of the thin, cambered wing for use in the low Re regime. For the sailplane, the mass budget is shown Table 5 and overall packaging needed for the 5 kg aircraft show in Table 6.

In order to provide adequate aerodynamic performance and controllability of the sailplane on Mars, several airfoils will be designed. Our previous studies[25,26] showed that the airfoil S5010-TOP24C[27,28] developed in our laboratory delivers the best performance at low Reynolds numbers in terms of the lift-to-drag ratio. This airfoil will be modified iteratively with the help of the distributed vortices method and program XFOIL[29].

The standard payload for this vehicle is an onboard HD camera and a transmitter of a total weight of $W_{CT} = 0.5$ kg. At the altitude of 500 m, processed images with a field of view of 260 by 195 m will be relayed to the control station on the ground providing continuous area coverage. The solar cells will give 12 hours of continues power supply to the on-board equipment and recharging batteries for overnight operations and its weight is $W_{SB} = 0.5$ kg. Noticeably, the sailplane will fly and operate fully autonomously. The weight of the autopilot, including IMU and computer, is approximately $W_{AP} = 0.3$ kg. Assuming the weight of the structure to be $W_{ST} = 1$ kg and a 1.5 kg margin total, the total mass of the sailplane should not exceed $m = 5$ kg. Note that the lightweight and inflatable design makes it possible to fit the craft inside the CubeSat.

**Table 1.** Mars Sailplane Mass Budget

| Major Subsystems | Mass |
|---|---|
| Science Camera | 0.4 kg |
| Inflatable Wing + Rudders, Frame | 1.0 kg |
| Communications | 0.7 kg |
| Computer & EPS | 0.3 kg |
| Battery | 0.5 kg |
| Attitude Determination & Control | 0.6 kg |
| System Margin (30 %) | 1.5 kg |
| Total | 5.0 kg |

**Table 2.** Mars Sailplane Packaging

| Major Subsystems | Mass |
|---|---|
| *Mars Sailplane Aircraft* | *5.0 kg* |
| Cold-Gas Propulsion | 3.0 kg |
| Nitrogen Generator | 3.7 kg |
| Astrotube Boom Deployer x 2 | 3.0 kg |
| Chassis | 2.0 kg |
| System Margin (30 %) | 7.3 kg |
| Total | 24.0 kg |

The equilibrium equation in the vertical direction for the level, steady flight can be written as follows

$$mg = L = 0.5 C_L \rho V^2 S \qquad (1)$$

Using this equation and assuming the max cruise speed of 100 m/s and the cruise lift coefficient of 0.248, the wing area is found, $S = 2.07$ m². With the imposed limitation on the size of the vehicle, specifically, the wing chord length should not exceed 0.36 m, the wingspan found to be 5.7 m. With a sea level viscosity of $\mu = 1.08 \cdot 10^{-5}$ N s/m², the Reynolds number can be found from the equation Re = 667·V. For the cruise speed $V = 100$ m/s, corresponding Reynolds number is 66,700. This value *falls* within low Reynolds number *range* (low Re is assumed to be less than $2 \cdot 10^5$). Design problems associated with low atmospheric density and low Reynolds number will be investigated theoretically and experimentally in the proposed project.

**MARS SAILPLANE DEPLOYMENT**

The Mars inflatable airplane concept will be deployed during EDL of a Mars-flagship rover/lander mission as the main EDL vehicle slows down to 100 m/s lateral velocity and achieves an altitude of 1.8 km (Figure 4). The Mars airplane package will be separate from the main EDL vehicle and use its onboard VACCO cold-gas propulsion [TRL 9][5] system to attain a 1 km separation distance. The gas generation system from MER/Pathfinder heritage [TRL 9] will be packaged into a 6U form-factor CubeSat attached to another 6U contain all the remaining components of the aircraft (Figure 5) [30] [32]. The gas generation system will produce non-toxic $N_2$ and pressurize the wing within 3 seconds [32]. Once the wing has pressurized. Telescopic booms developed by Oxford Space Systems [TRL 6] will expand to provide full structural support for the wing structure within about 10 seconds followed by disposal of gas generator and boom deployer. After 15 seconds, the airplane will use its wings to stay afloat and utilize the first hour to UV-cure (harden) the wings.

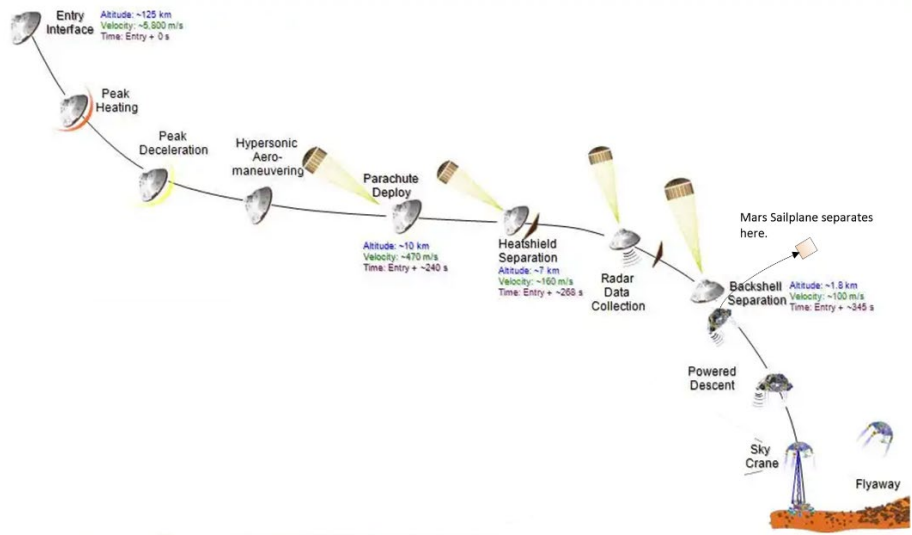

**Figure 4**. Entry, Descent and Landing Sequence for the Mars Science Laboratory/Mars 2020 Rover System. The Mars Sailplane Aircraft will separate from the EDL vehicle once it has slowed down to 100 m/s[31].

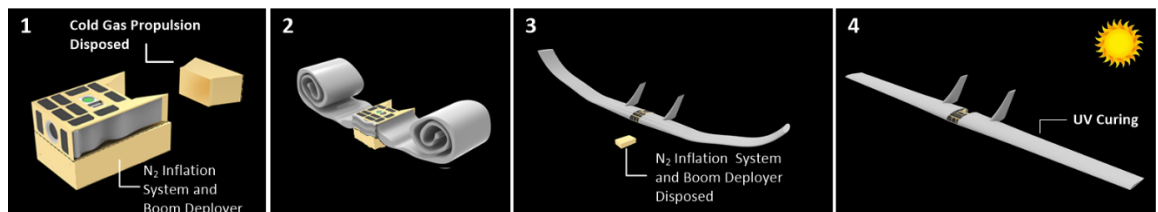

**Figure 5**. Mars sailplane deployment steps. First the cold-gas propulsion unit is used to achieve a 1 km separation distance from the EDL vehicle and then wing deployment begins (2-4).

Inflatable have been demonstrated during EDL on Mars Pathfinder and Mars MER Rovers[30]. They are also being developed for small satellites as communication antennas[32,33]. Goodyear and ILC Dover have developed inflatable aircraft. Goodyear's inflatable airplane was intended to carry a soldier[34]. ILC's inflatable wing is the most relevant to the concept presented (Figure 6) [35]. The wing is rolled up for deployment. Our proposed approach (Figure 5) for the wing will also to be roll-up the wings for stowage and using the inflation to pressure unroll and achieve pressure rigidization.

This will be followed by permanent rigidization through UV curing of the top side of the wings over a 40-minute period. The proposed Selig SU 5010 airfoil[27,28] is thick, for the 7,300-66,700 Reynold number, with very little camber (curvature) than makes the shape attainable. The wing contains a carbon fiber telescopic tube at the aerodynamic center. Oxford Space System's Astrotube has been shown to expand 2.5 m and hold a 0.3 kg load (mass of wing) on earth (Figure 6 right) [36]. It has been deployed to 3 m span in LEO making it TRL 6[36]. Additionally, the inflated wings may be filled with polyurethane foam to provide further rigidity and take 2 hours to cure.

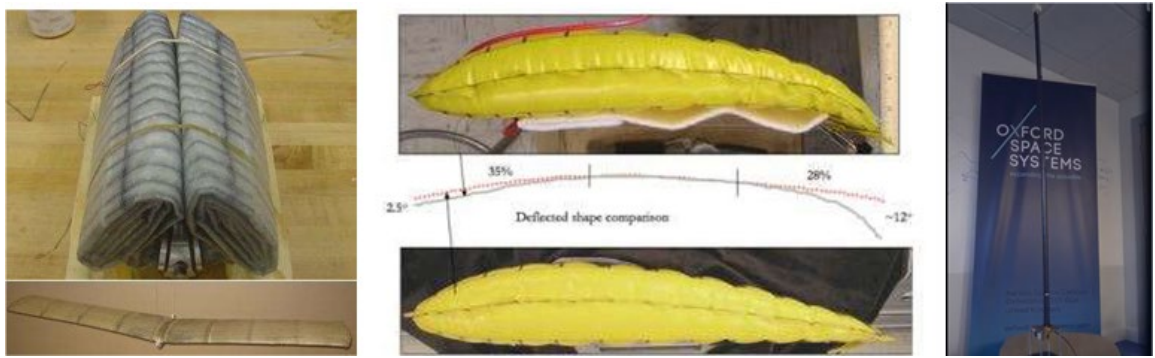

**Figure 6.** ILC Dover Inflatable wing stowed[35] (left) and inflated cross section[35] (center) and Astrotube (right) deployed at 2.5 m with a 300 g

We consider the advantage of an inflatable Mars aircraft vs. solid, folded rigid wing aircraft. Table 3 shows that folded wing aircraft would increase launch cost by 4 times and be too big in terms of mass to utilize the ballast mass on the flagship missions.

**Table 3.** Mass, Volume and Launch Cost for Mars Aircraft Concept.

|  | Mass [kg] | Height [m] | Width [m] | Length [m] | Launch Cost |
|---|---|---|---|---|---|
| Folded Rigid Wing | 100 | 0.4 | 0.4 | 6 | **$150 million** |
| Inflatable Wing | 24 | 0.24 | 0.24 | 5.7 | **$36 million** |

## DYNAMICS SOARING IN MARTIAN ATMOSPHERE

Dynamic soaring will be possible on Mars, thanks to the powerful numerical tools developed to model and simulate dynamic soaring on Earth. These very same models adapted to Mars conditions predicts dynamic soaring is possible (Figure 7). Dynamic soaring of sailplanes in the Earth's atmospheric boundary layer has been studied by using simulations and flight tests. Analysis of work-energy relationship of a sailplane in the wind is presented in[37]. The study showed that the energy neutral cycle depends on the maximum lift-to-drag ratio of the vehicle and the wind speed gradient.

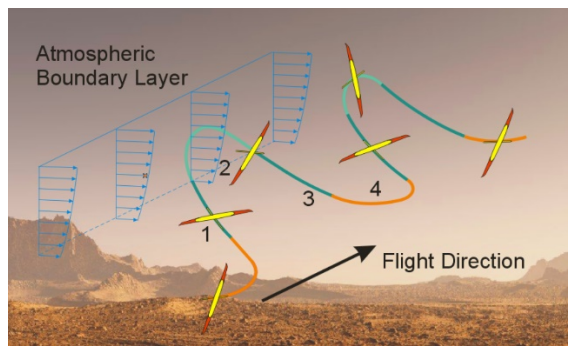

**Figure 7.** Dynamic soaring flight path (1 – windward climb, 2 – high-altitude turn, 3 – leeward descent, 4 – low altitude turn). Flight path cycles are utilized to permit perpetual flight.

For a continuous wind profile, the minimum gradient of the wing was calculated providing the neutral energy cycle.

Using trajectory optimization algorithm, the dynamic soaring maneuvers of unmanned aerial vehicles in atmospheric boundary layer were examined numerically for a number of different wind profiles[38]. Both 3DOF and 6DOF models were employed in[27] to demonstrate dynamic soaring with extreme climbs to high altitudes in high wind conditions. The dynamic soaring was modeled for the radio-controlled sailplane, which was then flown to demonstrate these maneuvers[39]. Flight tests on another sailplane revealed energy-conserving trajectories[40]. The previous studies of dynamic soaring on Earth proved its effectiveness in achieving long-endurance flights of weeks to months. Consider a flight path of a sailplane modeled as a point-mass, $m$, with three degrees of freedom. Figure 8 illustrates conventions for forces, angles, and velocities. There are three applied forces: lift, $L$, drag, $D$, and gravitational force, $mg$. The aerodynamic lift and drag are:

$$L = C_L 0.5 \rho V_a^2 S; \quad D = C_D 0.5 \rho V_a^2 S \tag{1}$$

and for the range of angle of attack, $\alpha$, below stall, the aerodynamic coefficients are

$$C_L = C_{L_\alpha} \alpha; \quad C_D = C_{D_{min}} + k C_L^2 \tag{2}$$

The three coefficients $C_{L_\alpha}, C_{D_{min}}$, and $k$ are functions of the sailplane configuration.

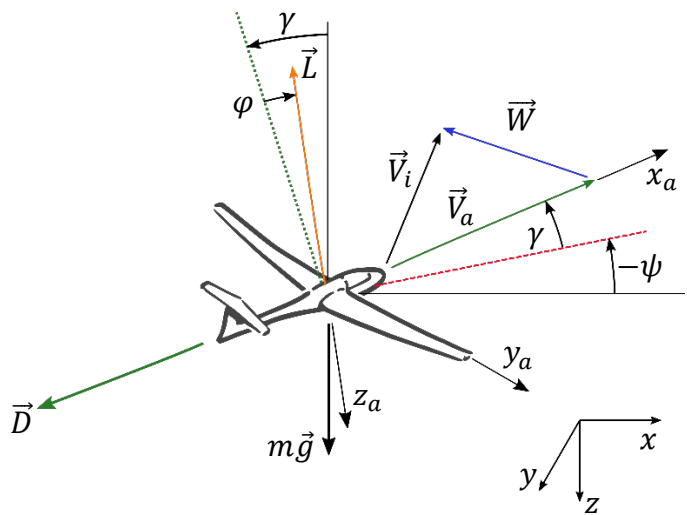

**Figure 8.** Conventions for forces, angles, and velocities.

The dynamic soaring problem is presented within the inertial reference frame $(x, y, z)$ and the frame $(x_a, y_a, z_a)$ attached to the sailplane. The transformation from the inertial to the aircraft reference frame is performed using three canonical rotations about the yaw angle, $\psi$, the pitch, $\gamma$, and the roll, $\varphi$, Euler angles. The components of the velocity vector of the vehicle relative to atmosphere are $\vec{V_a} = V_a[\cos(\psi)\cos(\gamma), \cos(\gamma)\sin(\psi), -\sin(\gamma)]^T$ and components of wind in the inertial frame are denoted as $\vec{W} = [W_x \ W_y \ W_z]^T$. Then corresponding kinematical equations are

$$\vec{V_i} = \vec{V_a} + \vec{W} = \begin{bmatrix} \dot{x} \\ \dot{y} \\ \dot{z} \end{bmatrix} = V_a \begin{bmatrix} \cos(\psi)\cos(\gamma) \\ \cos(\gamma)\sin(\psi) \\ -\sin(\gamma) \end{bmatrix} + \begin{bmatrix} W_x \\ W_y \\ W_z \end{bmatrix} \tag{3}$$

Equations of motion of the sailplane are obtained by applying the Newton's second law

$$m(\dot{V}_a + \dot{W}_x \cos(\psi) \cos(\gamma) + \dot{W}_y \cos(\gamma) \sin(\psi) - \dot{W}_z \sin(\gamma)) = -D + mg \sin(\gamma)$$
$$m(\dot{\gamma} V_a - \dot{W}_x \cos(\psi) \sin(\gamma) - \dot{W}_y \sin(\psi) \sin(\gamma) - \dot{W}_z) = L \cos(\varphi) + mg \cos(\gamma) \quad (4)$$
$$m(\dot{\psi} V_a \cos(\gamma) - \dot{W}_x \sin(\psi) + \dot{W}_y \cos(\psi)) = L \sin(\varphi)$$

Here in, the inertial wind rates are:

$$\dot{W}_x = \frac{\partial W_x}{\partial t} + \frac{\partial W_x}{\partial x}\dot{x} + \frac{\partial W_x}{\partial y}\dot{y} + \frac{\partial W_x}{\partial z}\dot{z}$$
$$\dot{W}_y = \frac{\partial W_y}{\partial t} + \frac{\partial W_y}{\partial x}\dot{x} + \frac{\partial W_y}{\partial y}\dot{y} + \frac{\partial W_y}{\partial z}\dot{z} \quad (5)$$
$$\dot{W}_z = \frac{\partial W_z}{\partial t} + \frac{\partial W_z}{\partial x}\dot{x} + \frac{\partial W_z}{\partial y}\dot{y} + \frac{\partial W_z}{\partial z}\dot{z}$$

Similarities in atmospheric boundary layers on Earth and Mars were revealed in the review[41] using numerical modeling and available empirical data. Based on these results, in the present study the assumption is made that Mars exhibits wind speed profiles that are comparable to those observed on Earth. Several previous studies on dynamic soaring near the Earth surface have been conducted using a simple linear wind velocity profile, where the wind speed increases linearly with altitude[38,42,43]. Simulations can be refined using models that more accurately describe the wind profiles, such as a logarithmic model[38,27] Combining Equations (3) and (4), the governing system of six first order differential equations can be written in the vector form

$$\dot{\vec{Y}} = f(\vec{Y}, \vec{u}), \quad (6)$$

where $\vec{Y} = [x, y, z, V_a, \gamma, \psi]^T$ and $\vec{u} = [C_L, \varphi]$ are the state vector the control vector, respectively. The time dependent control parameters $C_L$ and $\varphi$ affect the flight trajectory by changing either lift magnitude or heading. It is sufficient to control climb, descend, and turns of the sailplane. The constraints on the vehicle's motion are imposed on the lift coefficient by stall conditions

$$C_{L_{min}} \leq C_L \leq C_{L_{max}} \quad (7)$$

and by the limit load factor used in the aircraft design, $n$, presented in the form

$$C_L \leq n \frac{mg}{0.5 \rho V_a^2 S} \quad (8)$$

Allowing the lift coefficient to be negative eliminates the need for the constraint on the roll angle.

**Trajectory optimization**

There can be several special types of flight paths of aircraft, which employ a dynamic soaring, technique. These flights include typical cycles of climbing into a non-uniform wind, high-g turns and descending. When beginning and the end of each cycle are identified and this sequence of maneuvers is called a dynamic soaring cycle. During the cycle, the maximal total energy gain is found during a climb into the wind, where the increase in wind magnitude supplies additional kinetic energy and potential energy to the vehicle. At a speed close to the stall speed, the sailplane initiates a downwind turn followed by descend. At the end point, the state vector projection $\vec{Y}_p = [z, V_a, \gamma, \psi]^T$ is approximately equal to that at the initial point: $\vec{Y}_{pi} \approx \vec{Y}_{pe}$. The dynamic soaring cycle is referred to as a successful cycle, if the total energy of the sailplane, $E$, increases during the cycle, i.e., $E_e \geq E_i$. Successful cycles can be realized with the control parameters $C_L$ and $\varphi$, from which the optimal trajectory can be found. This representation formulates the optimization problem, in which the objective is to maximize the total energy along the flight path of the sailplane:

$$\max_{C_L, \varphi} \{E_{total} : \forall t\} \quad (9)$$

subjected to inequalities constraints:

$$\varphi_1 \leq \varphi \leq \varphi_2 \quad (10)$$

$$\psi_1 \leq \psi \leq \psi_2 \quad (11)$$

The solution to the optimization problem gives the optimal flight trajectory of the sailplane.

**Numerical results**

The purpose of this numerical study is to demonstrate the possibility of dynamic soaring trajectories in Mars's lower atmosphere. Equations of motion of a sailplane in dynamic soaring (9) along with the optimization problem have been solved numerically. Simulations were performed by employing the interior point optimizer method and software[44]. A simple Mars boundary layer model with linear wind profile was utilized. Specifically, this study considers only the variation of the horizontal wind with altitude in a steady state condition.

$$\dot{W}_x = \frac{dW_x}{dz}\dot{z} = -\frac{dW_x}{dz}V_a \sin(\gamma) \quad (12)$$

The wind profile was specified by the initial altitude, $z_i$, the wind speed, and the gradient $W_x|_{z=z_i} = W_{x0}$, and the wind speed gradient values were selected $dW_x/dz = [0.015, 0.019, 0.023]$. $|z_i| = 1.0$, $W_{x0} = 0$. Main characteristics of the sailplane and parameters of the mathematical model are given in Table 1.

**Table 4.** Sailplane model parameters for dynamic soaring.

| Parameter | Value | Parameter | Value | Parameter | Value |
|---|---|---|---|---|---|
| $M$ | 5.5 kg | $C_{L_\alpha}$ | - | $\varphi_1$ | $-60°$ |
| $S$ | 2.07m² | $C_{L_{max}}, |C_{L_{min}}|$ | 1.2 | $\varphi_1$ | $60°$ |
| $AR$ | 16 | $C_{D_{min}}$ | 0.015 | $\psi_1$ | $-180°$ |
| $N$ | 2 | $K$ | $1/\pi AR$ | $\psi_2$ | 0 |

Figure 9 shows three trajectories of dynamic soaring cycle corresponding to three values of wind gradient. The trajectories have a simple sinusoidal form with a maximum altitude reached at t = 60s. As the wind gradient $dW_x/dz$ increases the maximum altitude reached increases, the end state condition on altitude does not allow for overall altitude increase but in favorable conditions the energy gained at the end of the cycle shows the possibility of increased flight endurance.

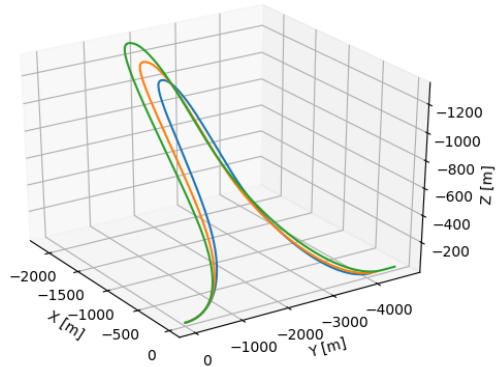

**Figure 9.** Trajectories of dynamic soaring cycles at different values of wind gradient (0.015 - blue, 0.019 - orange, 0.023 - green).

Success of this set of dynamic soaring trajectories is expressed with the total energy curves (Figure 10), the lowest wind gradient of 0.015 m/s$^2$ results in energy loss, whereas the two other cycles show sufficient energy gained from the wind environment to sustain perpetual advancing flight and result in a higher amplitude cycle. Results for $dW_x/dz = 0.019$ are close to the feasibility limit yet still shows evidence that dynamic soaring is possible in the lower Martian atmosphere under reasonable wind speeds ($dW_x/dz = 0.019 \equiv W_x(z\_i + 1000m) = 19 m/s$ ). It must be noted that each trajectory sees an equivalent amount of energy loss during the first 9 sec. of flight, this is attributed to the initial turn into the wind which doesn't constitute an energy gain phase.

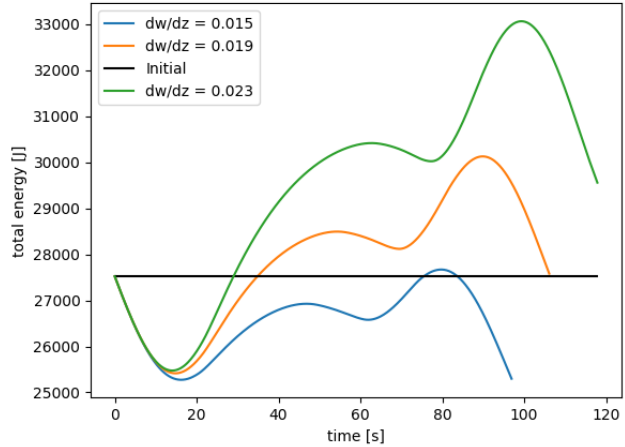

**Figure 10.** Total energy for three soaring cycles.

The turns described in the previous section are made evident by the relationship between the roll angle, yaw angle, and altitude in Figure 11. The roll angle reaches a limit of $\pi/2$ at the top of the trajectory where the sailplane needs to turn downwind in the smallest amount of time since this point in the trajectory is characterized by total energy loss, optimization of this maneuver will need to be conducted, to prevent such high roll angles that could not be favorable for the payload.

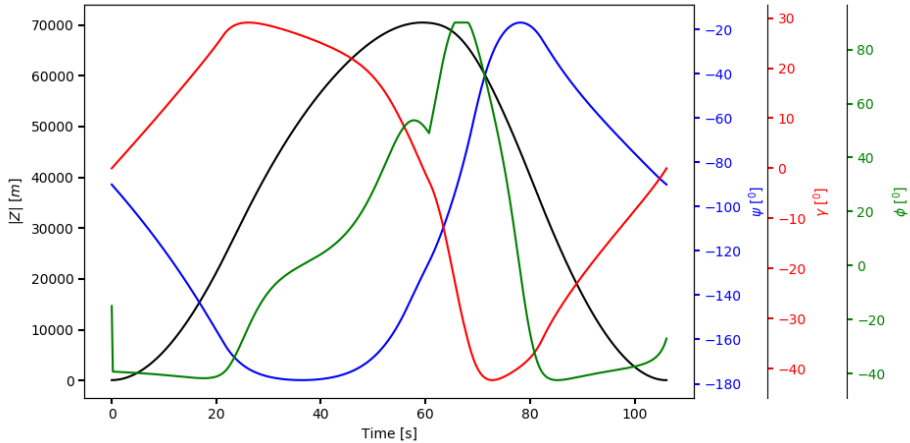

**Figure 11.** Flight angles and altitude over time at $dW_x/dz = 0.019$.

**CONCLUSIONS**

The Mars sailplane concept proposed here provides a whole new avenue for accelerating exploration of the Martian surface at a fraction of the cost of previous Mars airplane concepts. It will provide the highest resolution images of the Mars surface, better than what's possible with a satellite such as MRO and have longer flight-times, more than the Mars helicopter. This capability alone can shed light on several high-priority Planetary Science Decadal Survey questions, including the mystery behind Recurrent Slope Linaes and past/present habitability of Mars.  It could be used as a template for commercial and non-commercial partners to expand and provide a whole range of support services to NASA ranging from meteorology on Mars, to landing site hazard avoidance for a future Mars Sample Return Mission and future human mission to Mars. Importantly because the sailplane is a secondary payload better utilizing the 100+ kg ballast on a flagship mission, every second it operates provides invaluable data about Mars, advances CubeSat technology and sets a template rolling to explore other planets and moons with atmospheres such as Venus, Titan and the gas-giants.